\documentclass[a4paper]{article}
\usepackage{a4wide}
\usepackage{graphicx}
\usepackage{amsmath}
\usepackage{authblk}
\usepackage{subfigure}
\usepackage{tikz}

\usepackage{graphicx}
\graphicspath{{Figure/}}
\usepackage{bbm}

\usepackage{booktabs,multirow,multicol}
\usepackage{amsmath}
\usepackage{amsfonts} 
\usepackage{ulem}
\usepackage{amssymb}

\usepackage{amssymb}
\usepackage{changepage}

\usepackage{graphicx}
\graphicspath{{Figure/}}
\usepackage[labelfont=bf]{caption}

\usepackage{booktabs,multirow,multicol}
\usepackage{amsmath}
\usepackage{amsfonts}
\usepackage{ulem}
\usepackage{amssymb}

\usepackage{lineno}

\usepackage{bbm}
\usepackage{amsthm}
\usepackage{mathtools}
\usepackage{graphicx}
\usepackage{tikz}
\tikzstyle{mybox} = [draw=black, very thick, rectangle, rounded corners, inner ysep=5pt, inner xsep=5pt]

\theoremstyle{plain}
\newtheorem{thm}{Theorem}[section]

\theoremstyle{definition}
\newtheorem{defn}[thm]{Definition}

\theoremstyle{remark}

\title{Upscaling human activity data: an ecological perspective
}

\usepackage{lineno}
\author[a,b]{Anna Tovo \thanks{Corresponding author: anna.tovo@unipd.it}}
\author[a]{Samuele Stivanello \thanks{Corresponding author: stivanel@math.unipd.it}}
\author[b]{Amos Maritan}
\author[b,d]{Samir Suweis}
\author[c]{Stefano Favaro \thanks{Corresponding author: stefano.favaro@unito.it}}
\author[a,d]{Marco Formentin \thanks{Corresponding author: marco.formentin@unipd.it}}

\affil[a]{Dipartimento di Matematica \textquotedblleft Tullio Levi-Civita\textquotedblright, Universit\`a di Padova, Via Trieste 63, 35121 Padova, Italy.}
\affil[b]{Dipartimento di Fisica e Astronomia, \textquotedblleft Galileo Galilei\textquotedblright, Istituto Nazionale di Fisica Nucleare, Universit\`a di Padova, Via Marzolo 8, 35131 Padova, Italy.}
\affil[c]{Dipartimento di Scienze Economico-Sociali e Matematico-Statistiche, Universit\`a di Torino, Corso Unione Sovietica 218/bis, 10134 Torino, Italy.}
\affil[d]{Padova Neuroscience Center, Universit\`a di Padova, Via Orus 2/B, 35131  Padova, Italy.}

\begin{document}

\maketitle

\begin{abstract}
	In recent years we have witnessed an explosion of data collected for different human dynamics, from email communication to social networks activities. Extract useful information from these huge data sets represents a major challenge.  In the last decades, statistical regularities has been widely observed in human activities and various models have been proposed. Here we move from modeling to inference and propose a statistical framework capable to predict global features of human activities from local knowledge. We consider four data sets of human activities: email communication, Twitter posts, Wikipedia articles and Gutenberg books. From the statistics of local activities, such as sent emails per senders, post per hashtags and word occurrences collected in a small sample of
	the considered dataset, we infer global features, as the number of senders, hashtags and words at the global scale. Our estimates are robust and accurate with a small relative error. Moreover, we predict how abundance of a hashtag or of a word may change through scales. Thus, observing a small portion of tweets and the popularity of a given hashtag among them, we can estimate whether it will remain popular or not in the unseen part of the network. Our approach is grounded on statistical ecology as we discover inference of unseen human activity hallmarks can be mapped into the unseen species problem in biodiversity. 
	Our findings may have applications to different areas, from resource management in emails to collective attention monitoring in Twitter and to language learning process in word databases. 
\end{abstract}

\begin{figure}
	\caption{\textbf{From Ecology to Human Activities.} The figure depicts the correspondence between species/individuals in a natural ecosystem and users/sent emails, hashtags/posts, words/occurrences in each one of the four datasets considered in the paper. Once the proper correspondence is established, natural and artificial RSAs can both be well described by a negative binomial distribution. As exemplified in the last column, human activity RSA curves all display a fit with a negative value of $r$ in the interval $(-1,0)$, whereas natural ecosystems prefer $r>0$.
	}\label{fig:Fig1}
	\includegraphics[width=\textwidth]{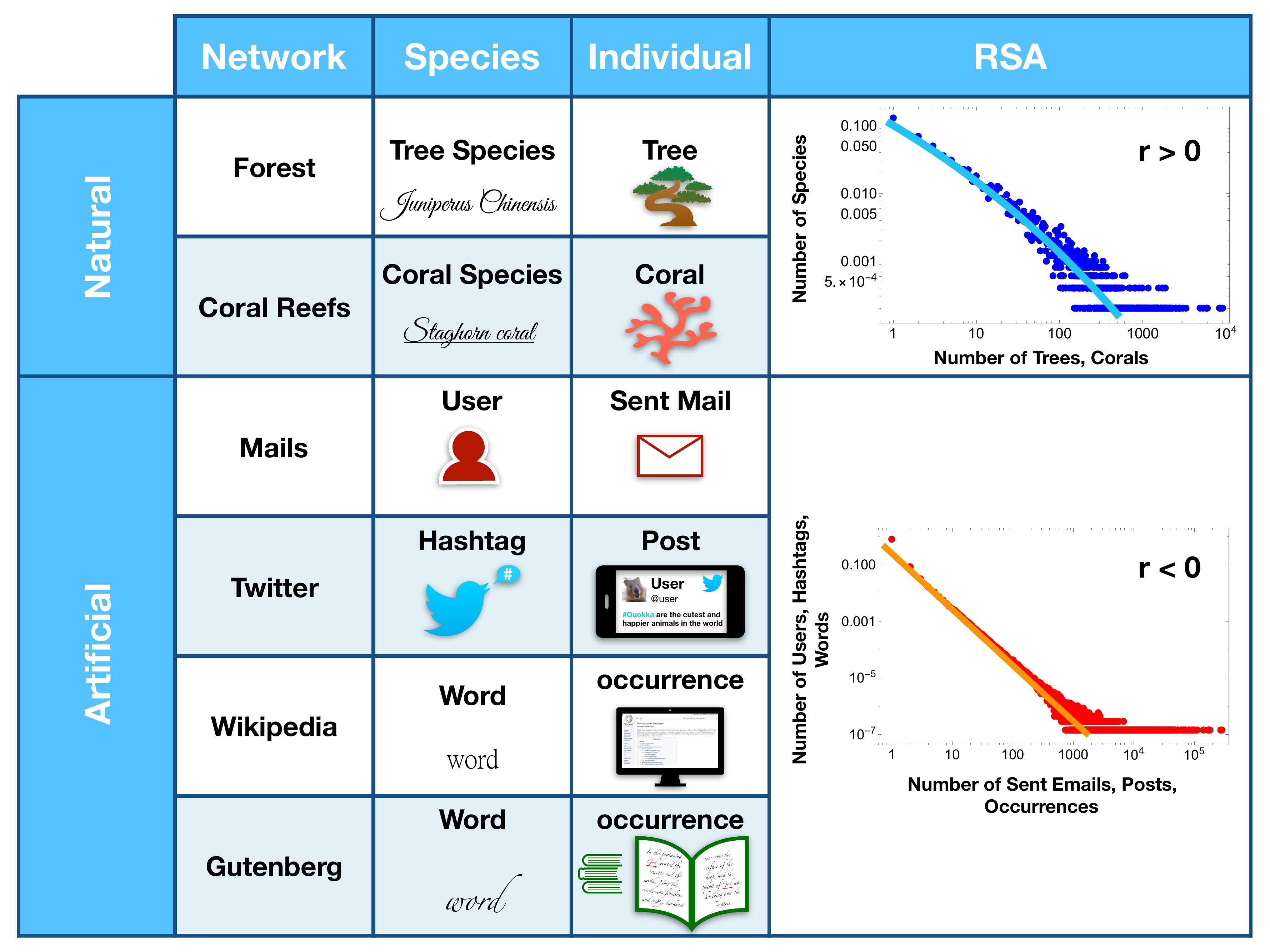}
\end{figure}

\begin{figure}
	\caption{\textbf{Theoretical framework.} Consider the email senders' network where each node is a sender and a directed link from node $A$ to node $B$ an email issued from user $A$ to user $B$. We set the identity of a sender to be the species and the sent emails to be the individuals of that species. For instance, if  $A$ has sent $n$ emails then the species $A$ has $n$ individuals. An observer sampling a fraction  $p$ of the sent emails, can partially recover the network (top-left) and the RSA curve at the local scale $p$ (bottom-left). Within our framework, this information suffices to infer the number of species  and  the RSA curve at the global scale $p=1$ (bottom-right). In terms of the network, the number of species corresponds to the number of users or nodes and the RSA gives the degree statistics. In this sense, our method reveals network features initially unknown to the observer, and pertaining to the whole community activity (top-right).  Moreover, we predict how the number of users increases with the number of links recorded, (i.e. the SAC curve in ecology), an information that may be used to optimize network design forecasting its growth. 
	}\label{fig:Fig3}
	\includegraphics[width=\textwidth]{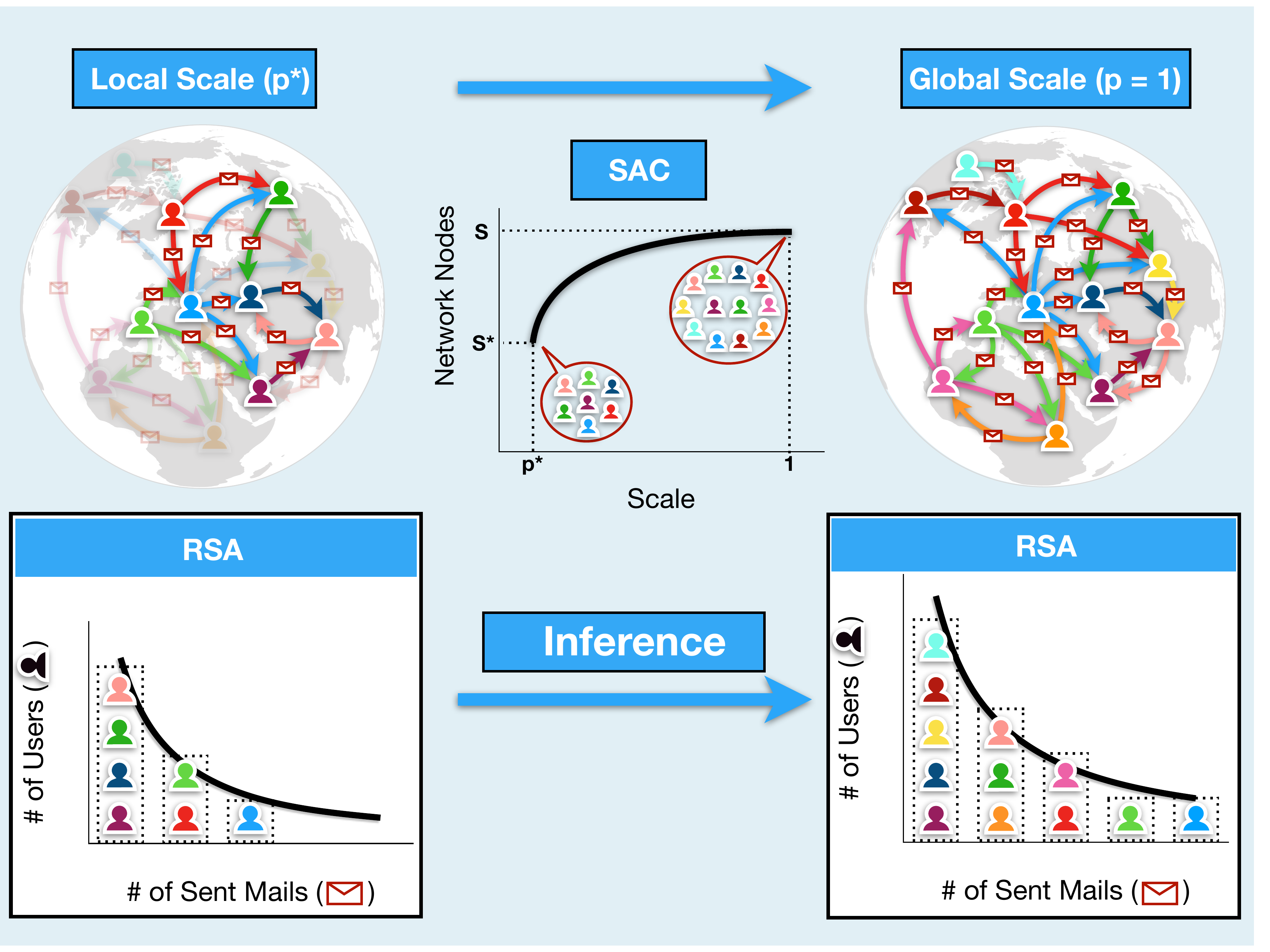}
\end{figure}

\begin{figure}
	\caption{\textbf{Universality and form invariance of the empirical RSAs.}  Empirical RSA curves at global scale ($p=1$) and local scale ($p=5\%$) are shown. RSA is scale-free in all the four datasets analyzed, with an heavy-tailed form maintained through different human activities and different scales. 
		RSA scale-invariance property allows for a successful implementation of our theoretical framework. In particular, our model predicts that the heavy-tail exponent $\alpha$ is related to the RSA clustering parameter $r$ via the relation $\alpha=1-r$ (see Supplementary Section S2.2). In each plot, for a visual inspection, we inserted a black line with slope $-\alpha=-1+\hat{r}$, where $\hat{r}$ have been obtained by fitting the local patterns at $p=5\%$ through a negative binomial (see also \tablename~\ref{tab:Tab1}). We can see that such lines describes very well the heavy-tail regime of the empirical RSAs at both local and global scale in all four cases. For the fitting curves and the predicted RSA patterns, see Supplementary Figure S1.}\label{fig:Fig2}
	\includegraphics[width=\textwidth]{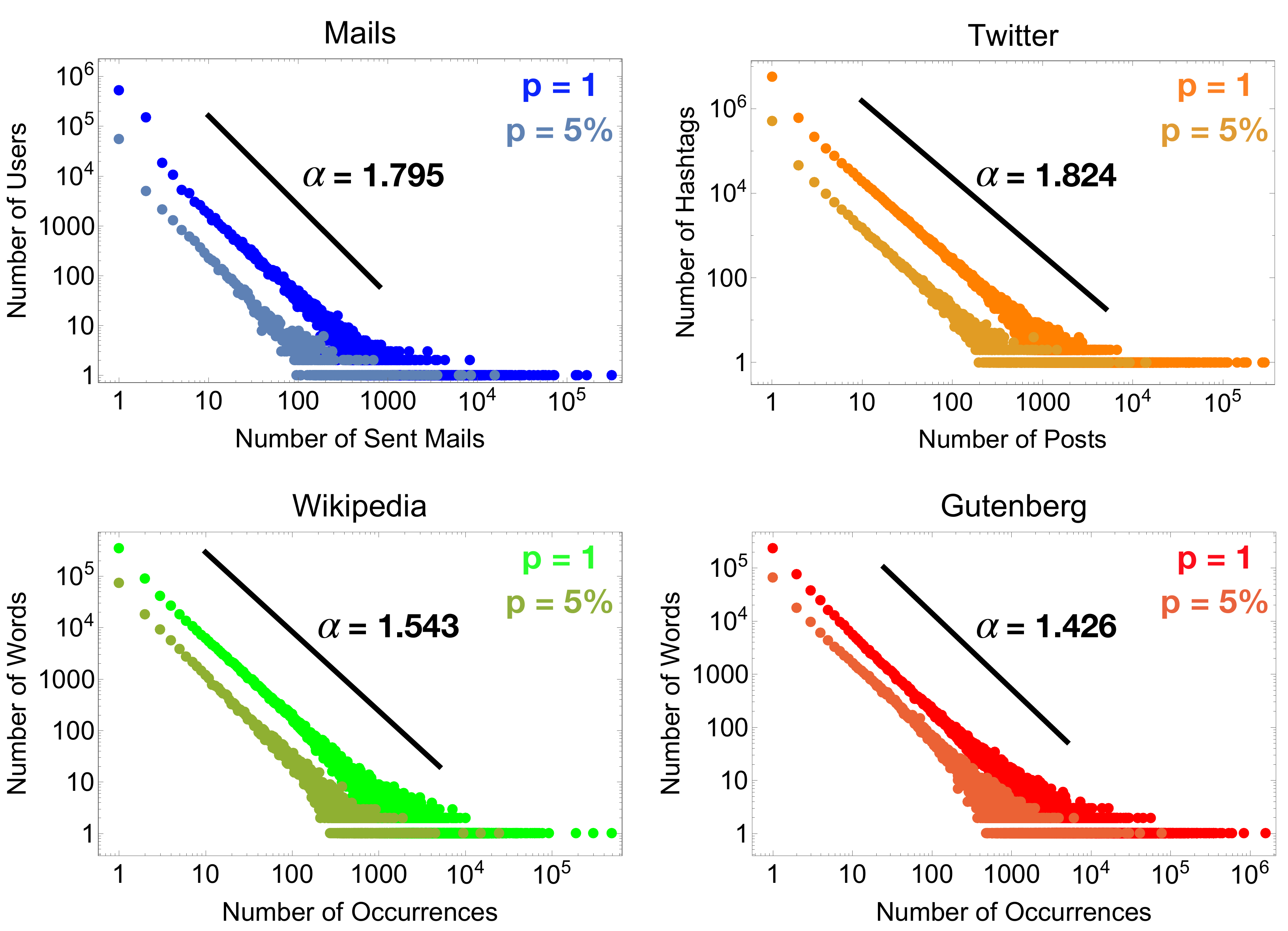}
\end{figure}

\bigskip
\noindent In ecology one of the most studied emerging pattern is the \textit{Relative Species Abundance} (RSA), that gives the fraction of species with the same number of individuals. Various ecosystems as tropical forests or coral reefs\cite{volkov2003neutral,volkov2007patterns,slik2015estimate,tovo2017upscaling}, despite their disparate locations and different evolutionary history, share a common shape of their empirical RSA (see \figurename~\ref{fig:Fig1}). Through years such an ubiquity encouraged  ecologists to develop several recipes to determine large scale RSA features from local information and diverse methods to estimate the number of species populating an ecosystem from knowing species abundances in a small portion of that ecosystem \cite{good1956number, harte2009biodiversity,chao2014species,slik2015estimate,orlitsky2016optimal}.
To figure the importance of such a challenge in ecology, let us recall that more than two-fifths of the number of worldwide trees can be found either in tropical or in subtropical forests, but only $\simeq 0.000067\%$ of species identities are known as very small portions of these forests have been sampled. For instance, only the $0.00016\%$ of trees of the Amazon forest has been surveyed \cite{ter2013hyperdominance,slik2015estimate,hubbell2015estimating}. \\
Recently\cite{tovo2017upscaling,tovo2019inferring}, it has been developed a rigorous statistical framework able to predict global scale biodiversity from scattered local plots. In the present paper we extend and generalize such a framework to human activity patterns and discuss the potential applications. In particular, we consider four datasets of human activities: e-mail communication \cite{formentin2014hidden,formentin2015new}, Twitter posts, Wikipedia articles and Gutenberg books \cite{tria2014dynamics}.
Statistical regularities in human dynamics have been widely observed in many different contexts and to understand such recurrent patterns a variety of models mainly grounded on statistical physics approaches have been proposed \cite{barabasi2005origin,alfi2007conference,malmgren2008poissonian,malmgren2009universality,loreto2011statistical,bagrow2011collective,loreto2012origin,yasseri2012circadian,torok2013opinions,gao2014quantifying,deville2016scaling,grauwin2017identifying,yasseri2017rapid}. \\
In the present work we move from modelling to inference. Our procedure is able to extract information at global scale starting from partial knowledge of the considered data sets. As as illustration of our ecological approach, of its potentiality and of the kind of results it can provide, we begin with e-mail communication.
The first step is to state what are the analogous of the species and individuals of a species in the email dataset. We consider the senders activity network where each node is a user and a directed link from node \textit{A} to node \textit{B} represents an email issued from user \textit{A} to user \textit{B}. We set the identity of a sender to label the species and the number of sent emails to be the individuals pertaining to a species. Thus, for instance, if user \textit{A} has sent $n$ emails we say that the species \textit{A} has $n$ individuals.  Suppose an observer have access to a small sample of sent emails, or, equivalently, to partial information on links and nodes of the email communication network. Our approach is capable to infer the number of nodes (i.e. the number of users) and the statistics of links of the whole network, thus revealing features of the dynamics previously unknown to the observer (see \figurename~\ref{fig:Fig3}).\\
In a similar manner, it is possible to set a correspondence between ecological species/individuals and human activities for the three other considered datasets. In the Twitter database, hashtags play the role of species and the number of different tweets containing a certain hashtag represents its population size. In Twitter and in social networks in general, popularity is known to be relevant, for instance, to manipulate mass opinion or to share information. One way to measure the popularity of a hashtag is to count the number of times the hashtag, appears in other users' tweets.  In our ecological interpretation, a hashtag represents a species, while the number of posts associated to it, gives the species' abundance. Therefore, in order to monitor the attention obtained by a hashtag, one should check which is position it reached along the RSA. 
In particular, if the species a given hashtag represents comes to be part of the right tail of the distribution, it means that the wished goal has been reached, since it constitutes one of the community dominant species, whereas if it comes to fall at the left tail of the RSA, it represents a hyper-rare species, thus not having received the wished attention.\\
For Wikipedia pages and Gutenberg book collection, throughout the paper we use the following settings: each word is a different species while its abundance is given by the number of occurrences of that word in the dataset (see \figurename~\ref{fig:Fig1}).\\ 
Once it has been defined  what are species and individuals of a species in each of the four human activities considered, our ecological perspective gives the following results:
\begin{itemize}
	\item \textbf{RSA universality and form invariance.} In each one of the mentioned dataset the RSA of the whole dataset (i.e. at scale $p=1$) turns out to be heavy tailed with exponent between -1.8 and -1.4 (see \figurename~\ref{fig:Fig2}). Moreover, the power law exponent is maintained at different scales.  With this we mean that when a portion of a database individuals are randomly sampled, the resulting RSA is still heavy tailed showing the same exponent as of the whole dataset (see \figurename~\ref{fig:Fig2}). We refer to this property as \textit{form invariance} of the RSA (see Supplementary Section S2.3).
	\item \textbf{Inference of unseen human activities.}  
	On the scale invariance property of the RSA we build a statistical framework  that gives robust and accurate estimates for the number of email senders, Twitter hashtags, Wikipedia pages and Gutenberg books from a random sample of sent mails, posts, word occurrences (see \tablename~\ref{tab:Tab1}).  Moreover, we can infer how abundance of a species may change through scales  (see \tablename~\ref{tab:TabPop}). This for example means that, observing a small portion of tweets and the popularity of a given hashtag among them, we can predict whether it will remain popular or not in the unseen part of the network. We refer to the inference of global quantities of interest from local  information as \textit{upscaling}. Finally, our framework predicts how the number of users/hashtags/words grows with the  activity (mails/posts/pages/books) recorded, which represents another well known pattern in ecological theory called the \textit{Species-Accumulation Curve} (SAC) (see \figurename~\ref{fig:Fig3}).
\end{itemize}

In the following we give the key steps of our upscaling framework. Denote with $N$ the population size and with $S$ the number of species (i.e. senders, hashtags, words) of the whole database. Given a scale $p^*\in(0,1)$, consider a random sample of size $p^*N$ in which we recover $S_{p^*}\leq S$ species. In the following we denote by $P(n|p^*)$ the fraction of species with $n$ individuals at scale $p^*$, i.e.  the sample RSA. We assume that, at the global scale $p=1$, $P(n|1)$ is proportional to a negative binomial distribution, $\mathcal{P}(n|r,\xi)$, with parameters $r\in(-1,+\infty)$ and $\xi\in(0,1)$: 
\begin{equation} \label{RSAglobal-MAIN}
	P(n|1) = c(r,\xi)\mathcal{P}(n|r,\xi) \mbox{\hspace{1.5cm} for } n \geq 1
\end{equation}
where the normalizing factor $c(r,\xi)=1/(1-(1-\xi)^r)$ takes into account that each of the $S$ species consists of at least one individual at the global scale.\\
RSAs given in \eqref{RSAglobal-MAIN} have the following features: 1) values of $r\in(-1,0)$ reflect in a heavy-tailed behavior of the RSAs, which well describes the observed patterns in human activities (see Supplementary Figure S1).  Indeed, the right tail of \eqref{RSAglobal-MAIN}  has the form $n^{1-r}\exp(n\log \xi)$ (see Supplementary Section S2.2) .  The exponent $\alpha=1-r$  matches very well with the empirical data (see also \figurename~\ref{fig:Fig2}). The exponential cutoff disappears in the limit $\xi\rightarrow 1$, for which  \eqref{RSAglobal-MAIN}  describes a pure power-law tail behavior.
2) Distribution \eqref{RSAglobal-MAIN} is \textit{form invariant}, meaning that the RSA $P(n|p)$ maintains the same functional form at different scales $p$ (see Supplementary Section 2.3), property observed in the empirical RSA of all the four databases (see \figurename~\ref{fig:Fig2}). In mathematical terms, the RSA at any scale $p$ is again proportional to a negative binomial with same $r$ and rescaled parameter \begin{equation}
	\xi_{p}=  p \xi / (1 - \xi (1 - p)).\label{eq:Eqxi}
\end{equation}
Properties 1) and 2) are the building blocks of our predictive statistical framework.\\
Our goal is to infer the total amount of species $S$ (senders, hashtags, words) in the complete database given the number of species $S_{p^*}$ observed in a sample at scale $p^*$ and their corresponding abundance (number of mails, posts, occurrences). From this limited information, we can construct the empirical values of the RSA, $P(n|p^*)$, and fit it to obtain the estimates $\hat{r}$ and $\hat{\xi}_{p^*}$ of the parameters that best capture the behavior of our data (henceforth we will denote with $\hat{\cdot}$ our estimation of any parameter $\cdot$). That, in turn, thanks to the form-invariance property, gives the value of the global parameter $\hat{\xi}$ via eq.~\eqref{eq:Eqxi}.\\
Observe that the probability that a given species present at $p=1$ is missing at $p<1$ corresponds to the fraction of non-observed species $(S-S_{p}) / S$. This value must be equal to $P(0|p)=1 - c(r,\xi)/c(r,\xi_{p})$, the probability for a species to have zero population in a sample of size $pN$ (see Supplementary section 2.4). Thus:
\begin{equation} \label{S-MAIN}
	\hat{S}\: \simeq \frac{S_{p^*}}{1-P(0|p^*)} \: \simeq \:  \frac{ 1 - (1 -\hat{\xi})^{\hat{r}} }{ 1 - (1 - \hat{\xi}_{p^*})^{\hat{r}} }S_{p^*} , 
\end{equation}
where the approximation is obtained by the definition of $c(r,\xi)$ and expressing $\hat{\xi}$ as a function of $\hat{\xi}_{p^*}$ by inverting eq.~(\ref{eq:Eqxi}).\\
To test the reliability of estimator (\ref{S-MAIN}), we extracted, from each dataset, ten sub-samples each covering a fraction $p^* = 5\%$ of the databases' individuals (sent emails, posted hashtags, occurrences of words), and we inferred the total number of species (email senders, posted hashtags in Twitter data and  different words in Wikipedia pages and Gutenberg books) from the empirical RSA constructed at $p^* = 5\%$. The average relative upscaling error is small in all four cases: about 0.1\% for sent Emails, 3\% for Twitter hashtags, 6\% for Wikipedia words and -2\% for Gutenberg words (See Table 1).  In \tablename~\ref{tab:Tab1} we report the average values of the fitted parameters together with the average relative percentage error between the predicted number of species, $\hat{S}$, and the true one, $S$.\\

The second innovation that we introduce in our work is a method to estimate the variation of popularity, a fundamental concept arising naturally when investigating human dynamics \cite{mestyan2013early,shen2014modeling,zhao2015seismic,yucesoy2016untangling,sinatra2016quantifying,jia2017quantifying}. Indeed, until now we studied the distribution of the abundances of the observed species at the local scale, but we estimated only the number of unseen species, disregarding of their abundances. Instead, abundance information is essential if one is interested, for example, in finding the most active users of the e-mail network or in the commonest words in a book or
in the popularity of a hashtag in Twitter database. In particular,  focusing on this last example, popularity of a hashtag is proportional to the number of posts containing it that come to circulate within the network thanks to other users’ tweets. This information is contained in the RSA pattern. Indeed, species with low population (i.e. hashtags posted a low number of times) are those positioned in the left side of the curve, whereas species with high abundances (i.e. hashtags posted a high number of times) are located in its right tail. Our goal now is to build an estimator for the change in popularity of hashtags from a portion $p^*$ of observed tweets to the remaining $1-p^*$ tweets.\\
Let us thus denote with $L$ a fixed threshold of posts above which we consider an hashtag popular at the sampled scale $p^*$ and let us indicate with $S_{p^*}(\geq L)$ the number of species having abundance at least $L$ in the surveyed collection of posts. We wish to check whether these (locally) popular species result to be popular also in the unseen fraction of the network, $1-p^*$. Let us then denote with $K$ the fixed popularity threshold at the unsurveyed scale. We are looking for an estimator of the number of species having abundance at least $K$ in the $1-p^*$ unseen part of the tweets, given that they have abundance at least $L$ at scale $p^*$. These species, which we denote with  $\hat{S}_{1-p^*}(\geq K | \geq L)$ are therefore globally popular within the network.\\
From our theoretical framework, we derive an estimator of such a quantity (See Supplementary Section S2.5). 
We define $S_{p^*}(l)$ the number of species having abundance exactly $l$ at scale $p^*$ and $S_{1-p^*}(k|l)$ the number of species having abundance exactly $k$ at scale $1-p^*$ given that they have abundance exactly $l$ at scale $p^*$. Then an estimator of $S_{1-p^*}(k|l)$ can be obtained via the following (see Supplementary Section S2.5 for details)):
\begin{equation}\label{eq:Spkl}
	\hat{S}_{1-p^*}(k | l)  = S_{p^*}(l) \cdot \dfrac{\binom{k+l}{l}p^{*l}(1-p^*)^k  \binom{k+l+\hat{r}-1}{k+l}\hat{\xi}^{k+l}(1-\hat{\xi})^{\hat{r}}}{\binom{l+\hat{r}-1}{l}\hat{\xi}_{p^*}^l(1-\hat{\xi}_{p^*})^{\hat{r}}}
\end{equation}
An estimator for $\hat{S}_{1-p^*}(\geq K | \geq L)$ can thus be obtained by summing up (\ref{eq:Spkl}) for all $k\geq K$ and for all $l\geq L$.
We tested the above estimator by fixing the (arbitrary) value of the threshold $L$ equal to 25 and varying the value of $K$ in the (arbitrary) range from 219 to 548 for ten sub-samples of Twitter database (for different choices of $L$ and $K$ see Supplementary Section S3.2). The average errors obtained in the predictions are displayed in Table~\ref{tab:TabPop}. For all the considered cases, we achieved very good estimates, with an average relative percentage error below 0.2\% in absolute value.\\

To conclude, we presented a statistical framework unifying upscaling in  ecology and human activities. We tested our method in four databases: email senders activity, Twitter hashtags, words in Wikipedia pages and Gutenberg books.  Once set the correspondence to what we consider species and individuals of a species, our approach reveals that the RSA is scale-free in each mentioned dataset with an heavy-tailed form maintained at different scales - with roughly the same exponent - through the different human activities considered (see \figurename~\ref{fig:Fig2}). This form-invariant property allows for a successful implementation of our predictive statistical framework. However, the heavy tail of the observed RSA cannot be captured by a standard negative binomial distribution with $r\in\mathbb{R}^+$. Nevertheless, such behaviours can be accommodated when allowing the clustering parameter $r$ to take negative values, $r\in(-1,0)$ (see Supplementary Section S2.2 and Supplementary Figure S1). This allows us to exploit form invariance to build estimator of unseen human activities from random samples.
In particular,  from the statistics of activities (sent emails per senders, post per hashtags, word occurrences) at local scale, we infer the number of species (number senders, hashtags, words) at the global scale.  Also we predict how popularity of species changes with scale. An issue of evident importance when thinking of social networks like Twitter. Finally, we compare our estimate with the true known value and in all the four databases considered the relative error is small (see Table 1, Table 2 and Supplementary Section S3.1). This result confirms the ability of the theoretical method to capture hidden feature of the human dynamics when only local information are available and pave the way for new applications in upscaling problems beyond statistical ecology.\\

{Our findings may have applications in different situations, spreading from resource management in emails to collective attention monitoring in Twitter and to language learning process in word databases. Let us see one example for each aforementioned context of how our framework could 
	help in decision making processes related to different aspects of social activity network.
	Let us start from the resource managing application. Suppose an internet/email provider starts a campaign to increase customers; for instance the provider wishes to double the number of subscribers. Now, in order to predict if more resources (e.g. number of server in the email example) are necessary to supply the newly entered subscribers, the provider needs to infer the total amount of activity bursting thanks to these new users. Our method provides a possible solution to this inference problem. 
	Indeed, by inverting eq. (\ref{S-MAIN}), which represents the well-known Species-Accumulation Curve in theoretical ecology, one obtains an analytical link between the total amount of activity (for instance number of sent emails) and the number of users. In particular, we see the activity does not grow linearly with the users, as one may naively guess. Thus, the information our framework provides on Species-Accumulation Curve,  may help the provider to decide how many further resources are needed for the expected number of new users. Clearly, this knowledge is useful either to avoid money waste in case no further resources are required, or to provide new structures/servers in advance in order to safely support the user activity and not to loose unsatisfied customers. Moreover, being aware of how many new structures are needed  also helps balance for profit their costs of installation, managing and maintenance with the price of subscriptions.\\
	A second application regards attention monitoring and information spreading. Nowadays social networks constitutes a fundamental source for spreading information and disinformation as well. They have being exploited to influence the mass opinion and attention in many different social context, from politics to economy \cite{margetts2015political}. It is enough to think about the influencer phenomenon arising in almost all social networks. In Twitter, popularity of a user may be read from the number of times a hashtag s/he initiated, appears in other users' tweets. In our ecological interpretation, hashtags represent a species, while the number of post associated to it, gives the species' abundance. Therefore, if the species s/he represents come to be part of the right tail of the RSA distribution, it constitutes one of the community dominant species and thus we can say s/he is popular, whereas if it comes to fall at the left tail of the RSA, it is a hyper-rare species, thus failing not having received the wished attention. However, in order to control someone's position within the global network, it is necessary to know which is the RSA at the whole community scale, a datum usually not provided by the social network manager organization. Twitter, for example, only releases information on the total number of tweets posted across time or, thanks to the Sample Tweets APIs or Decahose stream service, real-time random samples covering small percentages (up to 10\%) of the total tweets.  
	With this information, our framework offers the possibility to fully reconstruct the global RSA as well as to monitor how the number of popular hashtags scales from the monitored sample up to the whole activity network.
	This latter information may also be useful for governments or public administrations in general that want to communicate important news (health information, emergence procedures, elections etc...) to the citizens. In particular, our method allows to know the number of further tweets one eventually needs to effectively spread the information, allowing thus to undertake the proper measures (a bigger publicity campaign to obtain more followers, development of bot applications, etc.) to achieve the goal. 
	\\
	Finally, our theoretical framework may also be exploited in language learning process monitoring. For example, let us suppose to be learning a foreign speech. One may be interested in the number of books that are needed to be read in order to be sure to expand one's own vocabulary in order for it to cover a fixed percentage of all the speech words. The Species-Accumulation Curve emerging in this context thanks to our ecological correspondence between words/species and occurrences/abundances can thus been interpreted in a broader sense as a learning curve, where in the x-axis one has the total number of words encountered during the learning process (by dialogue  experience, frontal lectures or personal readings) and where in the y-axis one has the number of different words he manages to properly exploit in her/his speech.
}

\begin{table}
	\centering
	\caption{\textbf{Predicted relative errors.} Upscaling results for the number of species of the four analysed datasets from a local sample covering a fraction $p^*=5\%$ of the global database. For each database, we display the number of species (users, hashtags, words)  and individuals (sent mails, posts, occurrences) at the global scale, together with the fitted RSA parameters at the sampled scale and the relative percentage error between the true number of species and the one predicted by our framework. See Supplementary Figure S1 for fitting curves and predicted global patterns of the RSAs in the four cases.}
	\medskip
	\medskip
	\begin{tabular}{ccccc}
		\hline
		& Emails & Twitter & Wikipedia & Gutenberg\\
		\hline
		Species & $780,142$ & $6,972,453$ & $673, 872$ & $554,193$\\
		Individuals & $6,914,872$ & $34,696,973$ & $29,606,116$ & $126,289,661$ \\
		$\mathbf{r}$ & $-0.795$ & $  -0.824$ & $-0.543$ & $-0.426$\\
		$\mathbf{\xi_{p^*}}$ & $0.9999$ & $0.9991$ & $0.9985$ & $0.9997$\\
		Relative Error &  0.112 $\pm$ 0.385\% & 3.33 $\pm$ 0.17\% & 6.11 $\pm$ 0.118\% & -2.30 $\pm$ 0.23\%\\
		\hline
	\end{tabular}\label{tab:Tab1}
\end{table}

%

\begin{table}
	\centering
	\caption{\textbf{Percentage errors for popularity change predictions in Twitter database.} 
		For fixed $L=25$ and different values $K$ (first and second column),
		we estimated, from ten different Twitter sub-samples, the number of species having   abundance at least $K$ at the unobserved scale $1-p^*=95\%$ given that they have abundance at least $L$ at the sampled scale $p^*=5\%$ via estimator (\ref{eq:Spkl}).
		The average true number of species $S_{1-p^*}(\geq K| \geq L)$ and the average one predicted by our method among the ten sub-samples are displayed in the third and fourth columns. Finally, in the last two columns, we inserted the mean and the variance of the relative error obtained in the ten predictions. Similar results have been obtained for other values of $L$ (see Supplementary Table 2).}
	\medskip
	\medskip
	\begin{tabular}{cccccc}
		\hline
		$L$ & $K$ & $S_{1-p^*}(\geq K| \geq L)$ & $\hat{S}_{1-p^*}(\geq K| \geq L)$ & Relative Error & Variance \\
		\hline
		$25$ & $219$ & $5977$ & $5976.80$ & $0.0018131$ & $0.0000282$ \\
		$25$ & $329$ & $5943$ & $5950.31$ & $0.0448228$ & $0.01097890$ \\
		$25$ & $439$ & $5667$ & $5688.88$ & $0.0896268$ & $0.0609518$ \\
		$25$ & $548$ & $5064$ & $5055.71$ & $-0.1793290$ & $0.0877951$ \\
		\hline
	\end{tabular}\label{tab:TabPop}
\end{table}
\newpage
\clearpage
\newcommand{\beginsupplement}{%
	\renewcommand{\thetable}{S\arabic{table}}%
	\renewcommand{\thefigure}{S\arabic{figure}}%
	\renewcommand{\thesection}{S\arabic{section}}%
	\renewcommand{\thesubsection}{S\arabic{section}.\arabic{subsection}}%
}
\renewcommand{\figurename}{Supplementary Fig.}
\renewcommand{\tablename}{Supplementary Tab.}

\beginsupplement
\section*{Supporting Information}
\section{Datasets}
\noindent The databases concerning human activities analyzed in the study are four: Emails, Twitter, Wikipedia and Gutenberg. Here we give a brief description of the data. For further details, see Table 1 in the main article or \tablename~\ref{table-upscaling-3}.\\
In the following we will refer repeatedly to species and individuals. Although these are natural concepts in ecology, in this new context of human activities we need to state clearly what we treat as species and what we consider individuals. 
To start, consider the senders activity network where each node is a sender and a directed link from node \textit{A} to node \textit{B} represents an email issued from user \textit{A} to user \textit{B}. We set the identity of a sender to label the species and the number of sent emails to be the individuals pertaining to a species. 
Thus, for instance, if user \textit{A} has sent $n$ emails we say that the species \textit{A} has $n$ individuals.
In a similar manner, it is possible to set a correspondence between species/individuals and human activities for the three other considered datasets.
Throughout the paper we use the following settings. In the Twitter database, hashtags play the role of species and the number of different tweets containing a certain hashtag represents its population size. For Wikipedia pages and Gutenberg book collection, each word is a different species while its abundance is given by the number of occurrences of that word in the dataset.\\

\noindent\textbf{Emails}\\
\noindent This dataset is a collection of almost 7 millions emails, that corresponds to the activity of a Department of the University of Padova during two years: 2012 and 2013.
The collected data are in the form $\{$sender, receiver, timestamp$\}$. For our analysis, we select the first column of the table. Senders play the role of species, and each email is labelled with the name of the sender \cite{formentin2014hidden}.\\

\noindent\textbf{Wikipedia}\\
\noindent Our data represents all words contained in a collection of Wikipedia pages. 
We label each different word with a different number. Note that the same word always maintain its correspondence to the same number, regardless of the Wikipedia page it belongs \cite{tria2014dynamics}.\\

\noindent\textbf{Gutenberg}\\
\noindent Similar to Wikipedia, the dataset consists again of a collection of words belonging to some books.
We collected words and abundances in the same way as we did for Wikipedia \cite{tria2014dynamics}.\\

\noindent\textbf{Twitter}\\
\noindent Our dataset consists of a table where each row is of the form $\{$timestamp, hashtag, user$\}$. For our purposes, we select the second column of the table. Hashtags play the role of species, and their abundances correspond to the number of times they are posted. Dataset can be found in http://kreyon.net/waves-of-novelties/ .

\section{Theoretical framework} 

\subsection{Statistical model}
\noindent Once it has been defined what are species and individuals of a species in each of the four human activities considered, we can proceed in the explanation of our statistical model from an ecological perspective.\\
We denote with $N$ the total population size, and with $S$ the number of different species.\\
The \textit{Species Abundance Distribution} (SAD) at a subscale $p$ depicts the number of species in a subpopulation of size $pN$ that consist of a certain number $n$ of individuals. In the following we will quote as RSA its probability distribution, denoted by $P(n|p)$.\\
Consider now the whole system, i.e. the entire population. We assume that, at the global scale $p=1$, the RSA is proportional to a negative binomial distribution with parameters $r$ and $\xi$. It reads:
\begin{equation} \label{RSAglobal}
	P(n|1) = \displaystyle{c(r,\xi)\cdot \mathcal{P}(n|r,\xi)} \mbox{\hspace{1.5cm} for } n \geq 1
\end{equation}
where $\mathcal{P}(n|r,\xi)$ is the well known negative binomial density function with parameters $\xi$ and $r$, i.e.
$$
\mathcal{P}(n|r,\xi) = \binom{n+r-1}{n} \xi^n (1-\xi)^r
$$
and the normalizing factor $c(r,\xi)$ takes into account the fact that each of the existing $S$ species at the global scale consists of at least one individual:
$$
\displaystyle{c(r,\xi)  = \left[\sum_{n=1}^{\infty}  \binom{n+r-1}{n} \xi^n (1-\xi)^r \right]^{-1}  = \frac{1}{1 - (1-\xi)^r}}.
$$

\noindent Through the paper we always consider the generalized Negative Binomial distribution with  $\xi \in (0,1)$  and $r \in \mathbb{R}^+$ where the binomial coefficient is expressed by means of Gamma functions, i.e. $\binom{n+r-1}{n} = \frac{\Gamma(n+r)}{\Gamma(n+1) \Gamma (r)}$.\\
The reason why we chose to model the RSA  with a Negative Binomial  will be clear in few lines. For the moment, let us anticipate that Negative Binomial has two properties that are essential for the development of our estimators: it is form invariant (see Section S2.3) and, varying the values of $\xi$ and $r$, it can well describe different tail behavior from exponential to power-law (see Section S2.4).



\subsection{Power Law tails of $NB(r,\xi)$ with $r \in (-1,0)$} \label{r<0}
\noindent Negative binomial density function with parameters $\xi$ and $r>0$ results to capture very well empirical RSA patterns in tropical forests \cite{tovo2017upscaling,tovo2019inferring}. The observed RSAs in the analyzed human-activity databases, although displaying a similar universal character, do show a different behavior, being characterized by heavy tails which was not for RSA in tropical forests (see \figurename~\ref{RSA-invariant-3} and Figure 3 of the main text). These heavy tails of the observed RSAs cannot be captured by a standard negative binomial distribution with $r\in\mathbb{R}^+$. Nevertheless, they can be accommodated when allowing the clustering parameter $r$ to take negative values, $r\in(-1,0)$, thus enabling us to adapt and generalize the theoretical work of \cite{tovo2017upscaling} to portray regular statistics for human activities and to use activity information on local scale to predict hidden features of the human dynamics at the global scale.\\
We wish to show now that this extension of the parameter region reflects in a power-law behavior of the RSA's tail with an exponential cutoff, which well describes the observed patterns in human activities.
We point our that both the parameters intervene in the shape of the RSA, being $r$ responsible for the power-law tail with exponent $\alpha = 1-r$ and $\xi$ for the position of the exponential truncation of the distribution.
Note that, although this section is purely theoretical, the predicted exponent $\alpha = 1-r$ matches very well our findings when we empirically fit the data.\\
We start by considering our truncated negative binomial distribution of parameters $r$ and $\xi$:
\begin{equation} \label{y(n)}
	\displaystyle{ P(n) = c(r,\xi) \binom{n+r-1}{n} \xi^n (1-\xi)^r }
\end{equation}
The following theorem holds true \cite{walraevens2012stochastic,flajolet2008analytic}.
\begin{thm}\label{thm}
	Let $Y(z)$ be the generating function of a discrete random variable having probability mass function $P(\cdot)$ with dominant singularity $R_Y$. 
	Let $\beta \in \mathbb{R} \setminus \{0,1,2,...\}$. If for $z \rightarrow R_Y$
	\begin{equation} \label{hyp}
		Y(z) \sim c_Y \left(1-z/R_Y\right)^{\beta},
	\end{equation}
	then the distribution $P_Y(n)$ satisfies
	\begin{equation} \label{result}
		P(n) \sim \frac{c_Y n^{- \beta - 1} R_Y^{-n}}{\Gamma(-\beta)}\qquad\qquad\text{for } n \rightarrow \infty,
	\end{equation}
	where $\Gamma(\cdot)$ is the Gamma function.
\end{thm}
\noindent We wish to apply this theorem to our truncated negative binomial distribution.
Let us first recall that a singularity of a complex function is a point in the complex plane where the function is not analytic. Examples are poles, square-root branch points and branch cuts.\\ 
We now start by examining the probability generating function:
\begin{equation} \label{generating-function}
	Y(z) = \sum_{n=0}^{\infty} P(n) z^n
\end{equation}
Observe that $P(n)$ is given in (\ref{y(n)}), and that the normalizing factor $c(r,\xi)$ does not play any significant role. This is due to the fact that we wish to investigate the singularities of $Y(z)$ and thus the factor $c(r,\xi)$ does not affect the result. Moreover, the tail of a truncated negative binomial is exactly the same of a standard negative binomial, hence we simply disregard of the truncation and conduct the analysis for a standard negative binomial.\\
Since we aim at finding the lowest-norm singularity of the probability generating function $Y(z)$, we proceed with the computation by replacing the term $P(n)$ in (\ref{generating-function}) with its definition:
$$
\begin{array}{rcl}
Y(z) & = & \displaystyle{ \sum_{n=0}^{\infty} \binom{n+r-1}{n} \xi^n (1-\xi)^r z^n} \\
\: & \: & \: \\
\: & = & \displaystyle{\sum_{n=0}^{\infty} \binom{n+r-1}{n} (z\xi)^n (1-z\xi)^r \cdot \frac{(1-\xi)^r}{(1-z\xi)^r}} \\
\: & \: & \: \\
\: & = & \displaystyle{ \frac{(1-\xi)^r}{(1-z\xi)^r} \cdot \sum_{n=0}^{\infty} \binom{n+r-1}{n} (z\xi)^n (1-z\xi)^r}.
\end{array}
$$
For $z\xi < 1$, i.e. for $z< \frac{1}{\xi}$, the sum converges to 1 as we are summing over $\mathbb{N}$ the marginals of a standard negative binomial with parameters $r$ and $z\xi$.\\
Thus we are left with
$$
Y(z)  =  \displaystyle{ \frac{(1-\xi)^r}{(1-z\xi)^r}}  =  \displaystyle{c_Y (1-z\xi)^{-r}} 
$$
It turns out that $Y(z)$ has a singularity at $z=1/\xi$.\\
We now wish to express $Y(z)$ as in (\ref{hyp}) to apply the theorem. In our case:
$$
Y(z) = \displaystyle{c_Y  (1-z\xi)^{-r} } = \displaystyle{c_Y (1-z/R_Y)^{\beta}},
$$
where we set $\beta = -r$ and $R_Y = \frac{1}{\xi}$. Thus, Theorem (\ref{thm}) provides a characterization of the tails of the (truncated) negative binomial:
\begin{equation}
	\displaystyle{P(n) \: \sim\: \frac{c_Y n^{r - 1} \xi^{n}}{\Gamma(-\beta)} \:=\: \frac{c_Y n^{r - 1} e^{n \log(\xi)}}{\Gamma(-\beta)}, }\mbox{\hspace{1.5cm}} n >>1.
\end{equation}
Note that, since $\xi<1$ and $r-1<-1$, we have that both the exponential and the power-law approach zero when $n$ increase to infinity. Hence the distribution resembles a power-law until $n$ is of order $\ln(n) \cdot \frac{r-1}{\ln(\xi)}$. The cutoff thus depends both on $r$ and on $\xi$. In particular, the power-law range is greater for sharper slopes, i.e. for bigger absolute values of $r-1$, and for values of $\xi$ approaching 1.

\subsection{Scale Invariance of the RSA}
\noindent Zooming at a sub-scale $p$, i.e. considering a subpopulation of size $pN$, we will recover $S_p\leq S$ species. Note that $S_p$ may depend on which $pN$ individuals we select, i.e. different samples of the same size may lead to different values of $S_p$. We wish to derive the distribution of the local RSA $P(k|p)$ under the hypothesis of random sampling.\\
Under random sampling, it can be proven that, if the RSA at the global scale is distributed according to (\ref{RSAglobal}), then the local RSA at scale $p$ is again proportional to a negative binomial, with rescaled parameter $\xi_p$ and same $r$:
\begin{equation} \label{RSAlocal}
	P(k|p) = 
	\left\{
	\begin{array}{{ll}}
		c(r,\xi) \cdot \mathcal{P}(k|r,\xi_p) & \:\:\:k\geq1 \\
		\: & \: \\
		1 - c(r,\xi)/c(r,\xi_p) & \:\:\:k=0 
	\end{array}
	\right.
\end{equation}
with
\begin{equation} \label{xi-p}
	\displaystyle{\xi_p = \frac{p \xi}{1 - \xi (1 - p)}}.
\end{equation}
The fact that the RSA maintains the same functional form at different scales will be central in our framework. We will refer to this property as {\slshape scale-invariance}. We wish now to prove that this is indeed the case.\\
Suppose that a species consists of $n$ individuals among the whole population. Under random sampling, the conditional probability that the species has $k$ individuals at the sub-scale $p$, given that it has total abundance $n$ at the global scale, is given by a binomial distribution of parameters $n$ and $p$:
$$
\mathcal{P}_{binom}(k|n,p) = \binom{n}{k} p^k (1-p)^{n-k} \mbox{\hspace{3cm}} k=0,...,n
$$ 
and $\mathcal{P}_{binom}(k|n,p)=0$ if $k>n$. Let us now prove that the RSA at the local scale $P(k|p)$ is indeed distributed according to (\ref{RSAlocal}).\\
We start by noticing that, in order to compute the probability that a species in the subpopulation has abundance $k\geq1$, we need to condition on the fact that the species has abundance $n$ at the whole scale $p=1$, and then to sum over $n$, i.e. 
$$
\begin{array}{rcl}
\displaystyle{P(k|p)} & = & \displaystyle{\sum_{n \geq k} \mathcal{P}_{binom}(k|n,p) P(n|1)} \\
\: & \: & \: \\
\: & = & \displaystyle{ \sum_{n \geq k} \binom{n}{k} p^k (1-p)^{n-k} \cdot c(\xi,r) \binom{n+r-1}{n} \xi^n (1-\xi)^r }\\
\: & \: & \: \\
\: & = & \displaystyle{c(\xi,r) \binom{k+r-1}{k} \left( \frac{p\xi}{1- \xi(1-p) } \right)^{k} \left( \frac{1-xi}{1- \xi(1-p) } \right)^{r}} \\
\: & \: & \: \\
\: & = & \displaystyle{c(\xi,r)\binom{k+r-1}{k} \xi_p^k (1-\xi_p)^r} \\
\: & \: & \: \\
\: & = & \displaystyle{c(\xi,r) \cdot \mathcal{P}(k|r,\xi_p)}
\end{array}
$$
with $\xi_p$ given in (\ref{xi-p}). For $k=0$ we have
$$
\displaystyle{P(0|p) =  1- \sum_{k \geq 1}\mathcal{P}_{sub}(k|p) = 
	1-c(\xi,r)\sum_{k \geq 1}  \mathcal{P}(k|r,\xi_p) = 1- \frac{c(\xi,r)}{c(\xi_p,r)}}.
$$
Our method proceeds as follows: after fitting the parameters $\hat{\xi}_{p^*}$ and $\hat{r}$ from the empirical RSA observed at a local scale $p^*$, by inverting (\ref{xi-p}) we upscale them so to obtain an estimation of the global parameter $\hat{\xi}$ at $p=1$. 
The formula reads explicitly:
\begin{equation} \label{xi-global}
	\xi = \displaystyle{\frac{\xi_{p^*}}{p^* + \xi_{p^*}(1-p^*)}}.
\end{equation}
Note that this scale invariance holds between any two scales $q\leq p$. Indeed, from
$$
\displaystyle{\xi_p = \frac{p \xi}{1 - \xi (1 - p)}} \mbox{ \hspace{1cm} and \hspace{1cm} }\displaystyle{\xi_q = \frac{q \xi}{1 - \xi (1 - q)}}
$$
we obtain
$$
\begin{array}{rcl}
\displaystyle{\xi_q} & = & \displaystyle{\frac{q \xi}{1 - \xi (1 - q)} = \frac{q \frac{\xi_p}{p + \xi_p(1-p)}}{1 - \frac{\xi_p}{p + \xi_p(1-p)} (1 - q)} =  \frac{q \xi_p}{p  + \xi_p(1-p) -\xi_p(1 - q)} } \\
\: & \: & \: \\
\: & = & \displaystyle{\frac{q \xi_p}{p  - \xi_p(p - q)} = \frac{\frac{q}{p} \xi_p}{1 -\xi_p(1 - \frac{q}{p})}}.
\end{array}
$$
With the same argument, for any $q\geq p$ it holds
\begin{equation} \label{xi-q}
	\displaystyle{\xi_q = \frac{\xi_p}{\frac{p}{q} + \xi_p (1-\frac{p}{q})}}.
\end{equation}
Hence what really matters is the relative ratio of the two scales.\\
Our goal is now to estimate the global biodiversity of the community.

\subsection{Estimator for the total number of species and SAC}
\noindent We proceed now in the description of our procedure. Recall that our method uses only the information we can infer from a sub-sample at some scale $p^*$. Therefore, we only have information on the abundances of the $S_{p^*}$ species present in the surveyed area.
We now wish to determine the relationship between the total number of species $S$ in the entire population, i.e. at $p=1$, and the number of observed species at the sub-scale $p^*$.\\
Note that the probability that a species of the existing $S$ has null abundance at scale $p^*$ corresponds to the fraction of unsurveyed species. Hence we obtain
\begin{equation} \label{P(k=0|p)}
	P (k=0|p^*) \simeq \frac{S- S_{p^*}}{S}.
\end{equation}
Arranging the latter equation, we get a formula to predict the total number of species: 
\begin{equation} \label{S}
	\begin{array}{ccl}
		\hat{S} & \overset{\mbox{\tiny eq (\ref{P(k=0|p)})}}{=} & \displaystyle{ \frac{ S_{p^*} }{ 1 - P(k=0|p^*) } } \\
		\: & \: & \: \\
		\: & \overset{\mbox{\tiny eq (\ref{RSAlocal})}}{=} & \displaystyle{ S_{p^*} \frac{ 1 - (1 - \hat{\xi})^{\hat{r}} }{ 1 - (1 - \hat{\xi}_{p^*})^{\hat{r} } } } \\
		\: & \: & \: \\
		\: & \overset{\mbox{\tiny eq (\ref{xi-global})}}{=} & \displaystyle{ S_{p^*} \frac{ 1 - \left(1 - \displaystyle{\frac{\hat{\xi}_{p^*}}{p^* + \hat{\xi}_{p^*}(1-{p^*})}}\right)^{\hat{r} } }{ 1 - (1 - \hat{\xi}_{p^*})^{\hat{r}} } }
	\end{array}
\end{equation}
Thus we derived a formula to estimate the total number of species given a sub-sample at scale $p^*$.\\
Note that we can do more. By sub-sampling at sub-scales $q\leq p^*$ we can measure directly $S_q$. For any $q \in (p^*,1)$ we can apply the same chain of equations with some slight modification to estimate $\hat{S}_q$.
To be precise, for any $q \geq p^*$ we obtain
\begin{equation}
	\hat{S}_q = 
	\displaystyle{ S_{p^*} \frac{ 1 - \left(1 - \displaystyle{\frac{\hat{\xi}_{p^*}}{\frac{p^*}{q} + \hat{\xi}_{p^*} (1-\frac{p^*}{q})}}\right)^{\hat{r}} }{ 1 - (1 - \hat{\xi}_{p^*})^{\hat{r}} } } = 
	\displaystyle{ S_{p^*} \frac{ 1 - \left(\displaystyle{   \frac{p^* \left(1-\hat{\xi}_{p^*}\right)}{p^* + \hat{\xi}_{p^*} \left(q-p^*\right)}   }\right)^{\hat{r}} }{ 1 - (1 - \hat{\xi}_{p^*})^{\hat{r}} } }.
\end{equation}
Hence we obtained an explicit formula describing the behavior of the Species-Accumulation Curve (SAC) for every $q\leq1$.\\
Moreover we can express the RSA at the global scale by plugging the estimated parameters $\hat{\xi}$ and $\hat{r}$ into (\ref{RSAglobal}).

\subsection{Popularity and abundance variation through scales}
\noindent The second innovation that we are going to introduce in our work is a method to estimate the variation of popularity. Note that until now we studied the distribution of the abundances of the observed species at the local scale, but we estimated only the number of unseen species, disregarding of their abundances. \\
Before getting to that, we wish to recall our previous findings using a more detailed notation which turns out to be essential in the following.
\begin{defn}
	For every $s=1,...,S$, we indicate with $n_s^{p^*}, \: n_s^{1-p^*}$ the abundance of the species $s$ in the observed (resp. unobserved) fraction $p^*$ (resp. $1-p^*$) of the population.
\end{defn}
\begin{itemize}
	
	\item First, we need to introduce a statistics: \hspace{5mm} $\displaystyle{S_{p^*} = \sum_{s=1}^S \mathbbm{1}_{\{n_s^{p^*} > 0 \}}}$
	
	\item Let us compute the mean of the above statistics:
	\begin{equation*}
	\begin{split}
	 \mathbb{E} \left[ S_{p^*} \right] &= \mathbb{E} \left[ \sum_{s=1}^S \mathbbm{1}_{\{n_s^{p^*} > 0 \}} \right]
	= \sum_{s=1}^S \mathbb{E} \left[  \mathbbm{1}_{\{n_s^{p^*} > 0 \}} \right]
	= \sum_{s=1}^S \mathbb{P}\left(n_s^{p^*} > 0 \right)
	\\
	&\\
	&= S \cdot P \left(k > 0 | p^* \right) = S \cdot \left[ 1 - P \left(k = 0 | p^* \right) \right]
	\end{split}
	\end{equation*}
	\item Arranging the latter equation, we can isolate the quantity we are interested to estimate:
	\begin{equation} \label{Smean}
		\displaystyle{ S = \frac{\mathbb{E} \left[ S_{p^*} \right] }{1 - P \left(k = 0 | p^* \right)}}
	\end{equation}
	
	\item An estimator of $S$ is obtained replacing the mean $\mathbb{E} \left[ S_{p^*} \right] $ by the observable $\hat{S}_{p^*}$:
	\begin{equation} \label{Snew}
		\displaystyle{ \hat{S} = \frac{ \hat{S}_{p^*}}{1 - P \left(k = 0 | p^* \right)}}
	\end{equation}
	With no surprise, we recover the same result as in (\ref{S}). We wish to stress that  this new formulation allows us to push further our investigation, as we are going to show.
\end{itemize}
We wish now to apply the same procedure to different statistics.\\

\noindent Recall that we are sampling $S_{p^*}$ species at scale $p^*$ from a pool consisting of $N$ individuals spread into $S$ different species. If a species $s$ is not observed in the sample at scale $p^*$, we say that $s$ is a \textquotedblleft new" species. The meaning of this definition can be easily explained. If you imagine to further sample your population, you can pick individuals belonging to species already observed or you can discover indeed \textquotedblleft new" species.\\
Consider then the following statistics for the new species:
\begin{equation} \label{S1-p}
	S^{\scriptscriptstyle{\text{new}}}_{1-p^*} = \sum_{s=1}^S \mathbbm{1}_{\{n_s^{p^*} = 0, n_s^{1-p^*} > 0 \}}.
\end{equation}
The following chain of equality turns out to be meaningful in the following:
\begin{equation*}
\begin{split}
S^{\scriptscriptstyle{\text{new}}}_{1-p^*} &= \sum_{s=1}^S \mathbbm{1}_{\{n_s^{p^*} = 0, n_s^{1-p^*} > 0 \}}
= \sum_{s=1}^S \mathbbm{1}_{\{n_s^{p^*} = 0, n_s^{1} > 0 \}}\\
&
= \sum_{s=1}^S \mathbbm{1}_{\{n_s^{p^*} = 0 \}}
= \sum_{s=1}^S \left( 1 - \mathbbm{1}_{\{n_s^{p^*} > 0 \}} \right)
= S - S_{p^*}.
\end{split}
\end{equation*}
We can recover an estimator for  the \textquotedblleft new" species from the known estimator for $S$.\\
This remark seems trivial, and the chain of equation above appears redundant. Nevertheless, it is crucial for the development of our work. We stress that the statistics $S^{\scriptscriptstyle{\text{new}}}_{1-p^*}$ uses both the information at the sample scale $p^*$ and the information contained in the unseen fraction of the population $1-p^*$, whereas the statistics for $S_{p^*}$ only consider the observed individuals.\\
Given now the statistics (\ref{S1-p})
representing the number of unobserved species in the sample of size $p^*$, which are instead present in the remaining population of size $1-p^*$, We wish to recover an estimator for the new species. We thus compute the expected value of the statistics $S^{\scriptscriptstyle{\text{new}}}_{1-p^*}$:
\begin{equation*}
\begin{split}
\mathbb{E} \left[ S^{\scriptscriptstyle{\text{new}}}_{1-p^*} \right] &= \mathbb{E} \sum_{s=1}^S \mathbbm{1}_{\{n_s^{p^*} = 0, n_s^{1-p^*} > 0 \}} 
= S \cdot \mathbb{P} \left( n_s^{p^*} = 0, n_s^{1-p^*} > 0 \right)  \\
&
= S \cdot \mathbb{P} \left( n_s^{p^*} = 0, n_s^{1} > 0 \right)  
= S \cdot \underbrace{\mathbb{P} \left( n_s^{p^*} = 0\right) }_{\displaystyle{P \left(k = 0 | p^* \right)}}.
\end{split}
\end{equation*}
The expected value turns out to be a product of two factors: 
$P \left(k = 0 | p^* \right) = \mathbb{P}  (n_s^{p^*} = 0)$, which can be computed via (\ref{RSAlocal}), and $S$, a quantity we can estimate through $\hat{S} = \frac{  S_{p^*} }{1 - P \left(k = 0 | p^* \right)}$ as derived in (\ref{Snew}).
Hence we derived the following estimator:
$$
\hat{S}^{\scriptscriptstyle{\text{new}}}_{1-p^*} = \displaystyle{\frac{  S_{p^*} }{1 - P \left(k = 0 | p^* \right)} \cdot P \left(k = 0 | p^* \right)}
$$
This procedure captures the techniques that we wish to use to derive more estimators.\\
This turning point leads us to new statistics that consider also the popularity.\\

\noindent Let us start from the statistics:
\begin{equation}
	S^{\scriptscriptstyle{\text{new}}}_{1-p^*}(l) = \sum_{s=1}^S \mathbbm{1}_{\{n_s^{p^*} = 0, n_s^{1-p^*} = l \}}.
\end{equation}
Note that if we get an expression for $S^{\scriptscriptstyle{\text{new}}}_{1-p^*}(l)$, than we could easily extend the result to
$$
S^{\scriptscriptstyle{\text{new}}}_{1-p^*}(\geq L) = \sum_{l=L}^\cdot S^{\scriptscriptstyle{\text{new}}}_{1-p^*}(l).
$$
Moreover, results from the previous section can be included here, simply noticing that:
$$
S^{\scriptscriptstyle{\text{new}}}_{1-p^*} = S^{\scriptscriptstyle{\text{new}}}_{1-p}(\geq 1) =  \sum_{l=1}^\cdot S^{\scriptscriptstyle{\text{new}}}_{1-p^*}(l).
$$
We proceed as before by computing the expected value:
\begin{equation}
	\begin{array}{rcl}
		\mathbb{E} \left[ S^{\scriptscriptstyle{\text{new}}}_{1-p^*}(l) \right] &=& \mathbb{E} \left[ \sum_{s=1}^S \mathbbm{1}_{\{n_s^{p^*} = 0, n_s^{1-p^*} = l  \}} \right] \nonumber \\
		&=& S \cdot \mathbb{P} \left( n_s^{p^*} = 0, n_s^{1-p^*} = l \right)  \nonumber \\
		&=& S \cdot \mathbb{P} \left( n_s^{p^*} = 0, n_s^{1} = l \right)  \nonumber \\
		&=& S \cdot \underbrace{\mathbb{P} \left( n_s^{p^*} = 0 | n_s^{1} = l \right)}_{\displaystyle{Binomial (n_s^1,p^*)}} \underbrace{\mathbb{P} \left( n_s^{1} = l \right) }_{\displaystyle{P \left(l | 1 \right)}},
	\end{array}
\end{equation}
where we used, to pass from $n_s^{1-p^*}$ to $n_s^1$ in the third equality, the fact that  
$$\displaystyle { \mathbb{P} \left( n_s^{p^*} = x, n_s^{1-p^*} = y \right) = \mathbb{P} \left( n_s^{p^*} = x, n_s^{1} = x+y \right) }.$$\\
Let us note now the following facts:
\begin{itemize}
	\item $\mathbb{P} \left( n_s^{p^*} = 0 | n_s^{1} = l \right) = \displaystyle{(1-p^*)^l}$, from the sampling binomial distribution.
	\item $\mathbb{P} \left( n_s^{1} = l \right) = \displaystyle{P(l |1)}$ is given by (\ref{RSAglobal}).
	\item $S$ is unknown, and we need an estimator for it.
\end{itemize}
Again, we can use the results of the previous subsection to define $\displaystyle{ \hat{S} = \frac{  S_{p^*} }{1 - P \left(k = 0 | p^* \right)}}$
and hence to obtain
\begin{equation}
	\hat{S}^{\scriptscriptstyle{\text{new}}}_{1-p^*}(l) = \hat{S} \cdot (1-p^*)^l \cdot P(l | 1) = \displaystyle{ \frac{  S_{p^*} }{1 - P \left(k = 0 | p^* \right)}} \cdot (1-p^*)^l \cdot P(l | 1), 
\end{equation}
which is the estimator for the new species with abundance $l$.\\
Thus, as a first partial result, we obtained an estimator for the popularity of the new species.\\

\noindent Let us now consider the statistics:
\begin{equation}
	S_{1-p^*}(l \rightarrow k) = \sum_{s=1}^S \mathbbm{1}_{\{n_s^{p^*} = l, n_s^{1-p^*} = k \}},
\end{equation}
which represents the number of species having contemporarily abundance $l$ at the observed scale $p^*$ and abundance $k$ at the unobserved scale $1-p^*$.
Note that we can compute also intervals of abundances by summing on different values of $l$ and $k$.
We proceed by computing the expected value:
\begin{equation}
	\begin{array}{rcl}
		\mathbb{E} \left[ S_{1-p^*}(l \rightarrow k) \right] &=& \mathbb{E} \left[ \sum_{s=1}^S \mathbbm{1}_{\{n_s^{p^*} = l, n_s^{1-p^*} = k  \}} \right] \nonumber \\
		&=& S \cdot \mathbb{P} \left( n_s^{p^*} = l, n_s^{1-p^*} = k \right)  \nonumber \\
		&=& S \cdot \mathbb{P} \left( n_s^{p^*} = l, n_s^{1} = k+l \right)  \nonumber \\
		&=& S \cdot \underbrace{\mathbb{P} \left( n_s^{p^*} = l | n_s^{1} = k+l \right)}_{Binomial(n_s^1,p^*)} \underbrace{\mathbb{P} \left( n_s^{1} = k+l \right) }_{\displaystyle{P(k+l | 1)}}.
	\end{array}
\end{equation}
Now we have the following:
\begin{itemize}
	\item $\mathbb{P} \left( n_s^{p^*} = l | n_s^{1} = k+l \right) = \displaystyle{\binom{k+l}{l}p^{*l}(1-p^*)^k}$, from the sampling binomial distribution;
	\item $\mathbb{P} \left( n_s^{1} = k+l \right) = \displaystyle{P(k+l | 1)} = \displaystyle{c(r,\xi)\binom{k+l+r-1}{k+l}\xi^{k+l}(1-\xi)^r}$;
	\item $S$ is unknown. However, we can estimate it via $\displaystyle{ \hat{S} = \frac{  S_{p^*} }{1 - P \left(k = 0 | p^* \right)}}.$
\end{itemize}
Hence we obtained
\begin{equation}
	\begin{array}{rcl}
		\hat{S}_{1-p^*}(l\rightarrow k) & = &
		\hat{S} \cdot \mathbb{P} \left( n_s^{p^*}= l | n_s^{1} = k+l \right) \cdot P(k+l |1) \nonumber \\
		& = & 
		\displaystyle{ \frac{  S_{p^*} }{1 - P(0|p^*)}} \cdot \displaystyle{\binom{k+l}{l}p^{*l}(1-p^*)^k} \cdot \displaystyle{c(r,\hat{\xi})\binom{k+l+\hat{r}-1}{k+l}\hat{\xi}^{k+l}(1-\hat{\xi})^{\hat{r}}}.
	\end{array}
\end{equation}
Estimator $\hat{S}_{1-p^*}(l \rightarrow k)$ above infers the number of species with abundance
$l$ at the observed scale $p^*$ and abundance $k$ at the unobserved scale $1-p^*$. 
Note that this estimator is independent of the the number of species with abundance
$l$ at scale $p^*$; indeed, we are using the sample at scale $p^*$ only to estimate the parameters $\xi_{p^*}$ and $r$, which we need to predict $\hat{S}$. Hence we are only using partial information at the local scale.\\

\noindent We wish now to take into account the information about the number of species with abundance $l$ at the surveyed scale, $S_{p^*}(l)$.\\
In particular, we are looking for an estimator of the species with abundance $k$ in the unobserved fraction $1-p^*$ of the population, given that they have abundance $l$ in the sample at observed scale $p^*$.\\
We thus define $S_{p^*}(l) := \sum_{s=1}^S \mathbbm{1}_{\{n_s^{p^*} = l \}}$.\\
In the following we will need to use quantities of the type $\displaystyle{\mathbb{P} ( n_s^{1-p^*} = k | n_s^{p^*} = l )}$. \\
Using Bayes' theorem, we obtain
$$
\begin{array}{rcl}
\mathbb{P} ( n_s^{1-p^*} = k | n_s^{p^*} = l ) & = & \mathbb{P} ( n_s^{1} -n_s^{p^*}= k | n_s^{p^*} = l )\\
& = & \mathbb{P} (n_s^{1} -l= k | n_s^{p^*} = l ) \\
& = & \mathbb{P} ( n_s^{1} = k+l | n_s^{p^*} = l) \\
& = & \displaystyle{   \frac{\mathbb{P}( n_s^{p^*} = l | n_s^{1} = k+l)   \mathbb{P} ( n_s^{1} = k+l ) }{\mathbb{P} ( n_s^{p^*} = l  ) }}.
\end{array}
$$
Note that we all the probabilities appearing in the latter formula are known, since:
\begin{itemize}
	\item $\mathbb{P} \left( n_s^{p^*} = l | n_s^{1} = k+l \right) = \displaystyle{\binom{k+l}{l}p^{*l}(1-p^*)^k}$ is the sampling binomial distribution;
	\item $\mathbb{P} \left( n_s^{1} = k+l \right)
	= \displaystyle{P(k+l |1)} = \displaystyle{c(r,\xi)\binom{k+l+r-1}{k+l}\xi^{k+l}(1-\xi)^r}$ is the global truncated negative binomial distribution as in (\ref{RSAglobal}) of parameters $r$ and $\xi$;
	\item $\mathbb{P} \left( n_s^{p^*} = l \right) = \displaystyle{P(l | p^*)} = \displaystyle{c(r,\xi)\binom{l+r-1}{l}\xi_p^{*l}(1-\xi_{p^*})^r}$ is again a truncated negative binomial with rescaled parameter $\xi_p$ as in (\ref{RSAlocal}).
\end{itemize}
Let us now retrace the same steps as for $\hat{S}_{1-p^*}(l \rightarrow k)$ for the conditional estimator $\hat{S}_{1-p^*}(k|l)$.
We start from the statistics
$$
S_{1-p^*}(k | l) = \displaystyle{ \sum_{s=1}^S \mathbbm{1}_{\{n_s^{p^*} = l \}} \mathbbm{1}_{\{ n_s^{1-p^*} = k,n_s^{p^*} = l \}}  = \sum_{s=1}^{S_{p^*}(l)} \mathbbm{1}_{\{n_s^{1-p^*} = k | n_s^{p^*} = l\}}   }
$$
We proceed by computing the expected value
$$
\mathbb{E} \left[ S_{1-p^*}(k | l) \right] =  
\displaystyle{ S_{p^*}(l) \cdot \mathbb{P} \left( n_s^{1-p^*} = k | n_s^{p^*} = l \right) } 
=
\displaystyle{ S_{p^*}(l) \cdot \frac{\mathbb{P} \left( n_s^{p^*} = l | n_s^{1} = k+l \right) \mathbb{P} \left( n_s^{1} = k+l \right)}{\mathbb{P} \left( n_s^{p^*} = l \right)} }.
$$
Note that empirically $\mathbb{P} \left( n_s^{p^*} = l \right) = S_{p^*}(l) / S $ so that we can recover $\mathbb{E} \left[ S_{1-p^*}(l \rightarrow k) \right].$\\
Let us now insert into the above formula the probabilities computed by using the fitted parameters:
$$
\hat{S}_{1-p^*}(k | l)  = S_{p^*}(l) \cdot \frac{\displaystyle{   \binom{k+l}{l}p^{*l}(1-p^*)^k} \cdot \displaystyle{\binom{k+l+\hat{r}-1}{k+l}\hat{\xi}^{k+l}(1-\hat{\xi})^{\hat{r}}}}{\displaystyle{\binom{l+\hat{r}-1}{l}\hat{\xi}_{p^*}^{l}(1-\hat{\xi}_{p^*})^{\hat{r}}}}, 
$$
where the terms $c(r,\hat{\xi})$ in the numerator has cancelled out with the one at the denominator.\\
Estimator $\hat{S}_{1-p}(k | l)$ is theoretically unbiased.\\ 
Note that, again, we can pass from punctual estimation to cumulative ones, by the summing up over all values $l$ and $k$ above fixed thresholds $L$ and $K$, respectively:
\begin{equation}\label{eq:CondEst}
	\hat{S}_{1-p^*}(\geq K | \geq L) =\sum_{l \geq L} \sum_{k\geq K} \hat{S}_{1-p^*}(k | l)
\end{equation}
and this is exactly the estimator we are going to test in our database.


\section{Additional results and figures}

\noindent In this section we collect results not presented in the main text.
\begin{figure}[h]	
	\includegraphics[width=\textwidth]{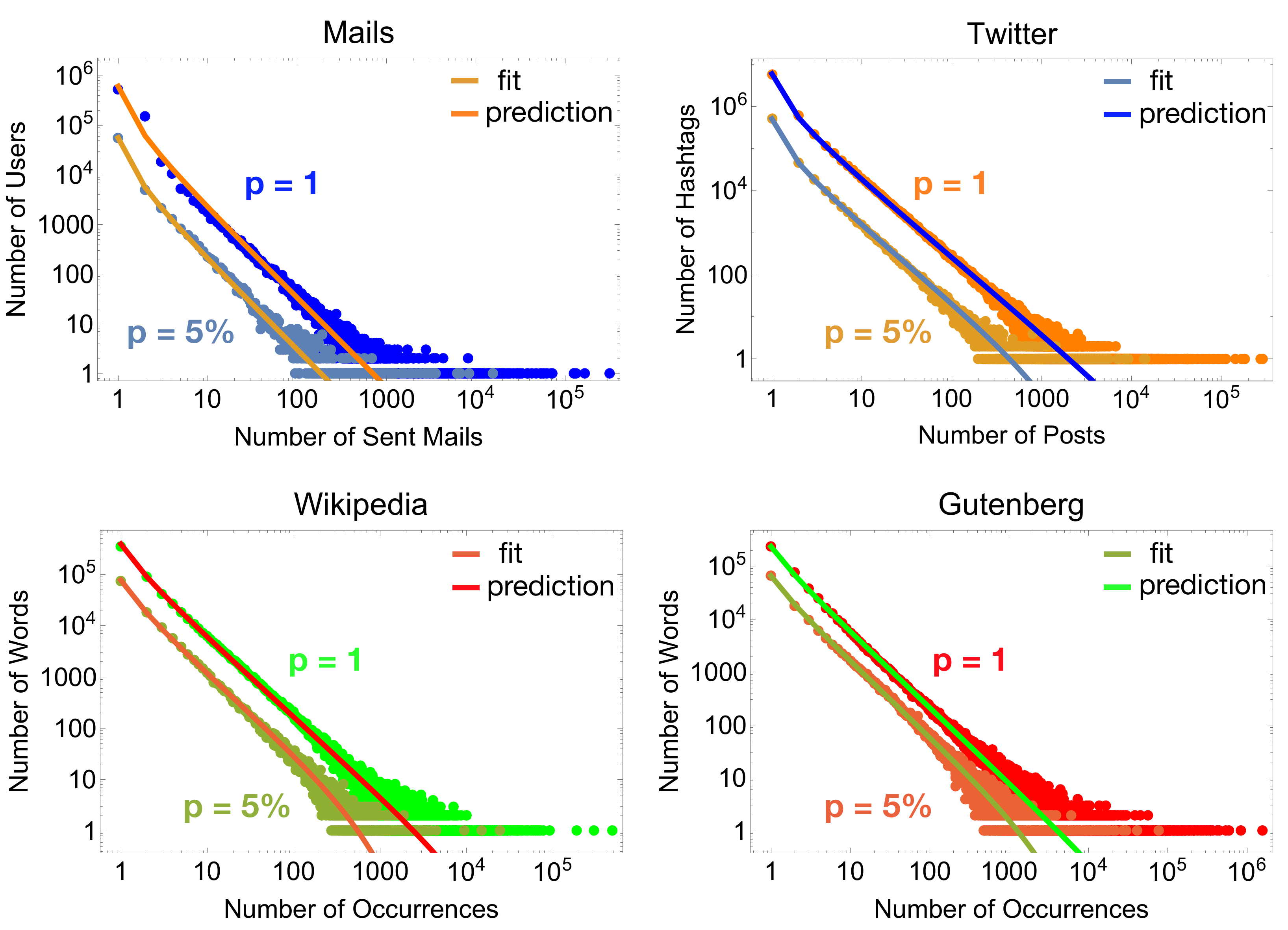}
	\caption{\textbf{Best-fit and upscaling predicted pattern from sample scale $p^*=5\%$.}  Empirical RSA curves at global scale ($p=1$) and local scale ($p^*=5\%$) are shown. In each panel, coloured lines over the local RSA represent the  distribution obtained via a best-fit of the empirical pattern with a negative binomial with $r\in(0,1)$. Lines over the global RSA represent our prediction for the RSA at the  global scale obtained via our upscaling equations for both the parameters and the biodiversity.}\label{BetsFit}
\end{figure}

\subsection{Upscaling results from sample scale $p^*=3\%$}
\begin{figure}[h]	
	\includegraphics[width=\textwidth]{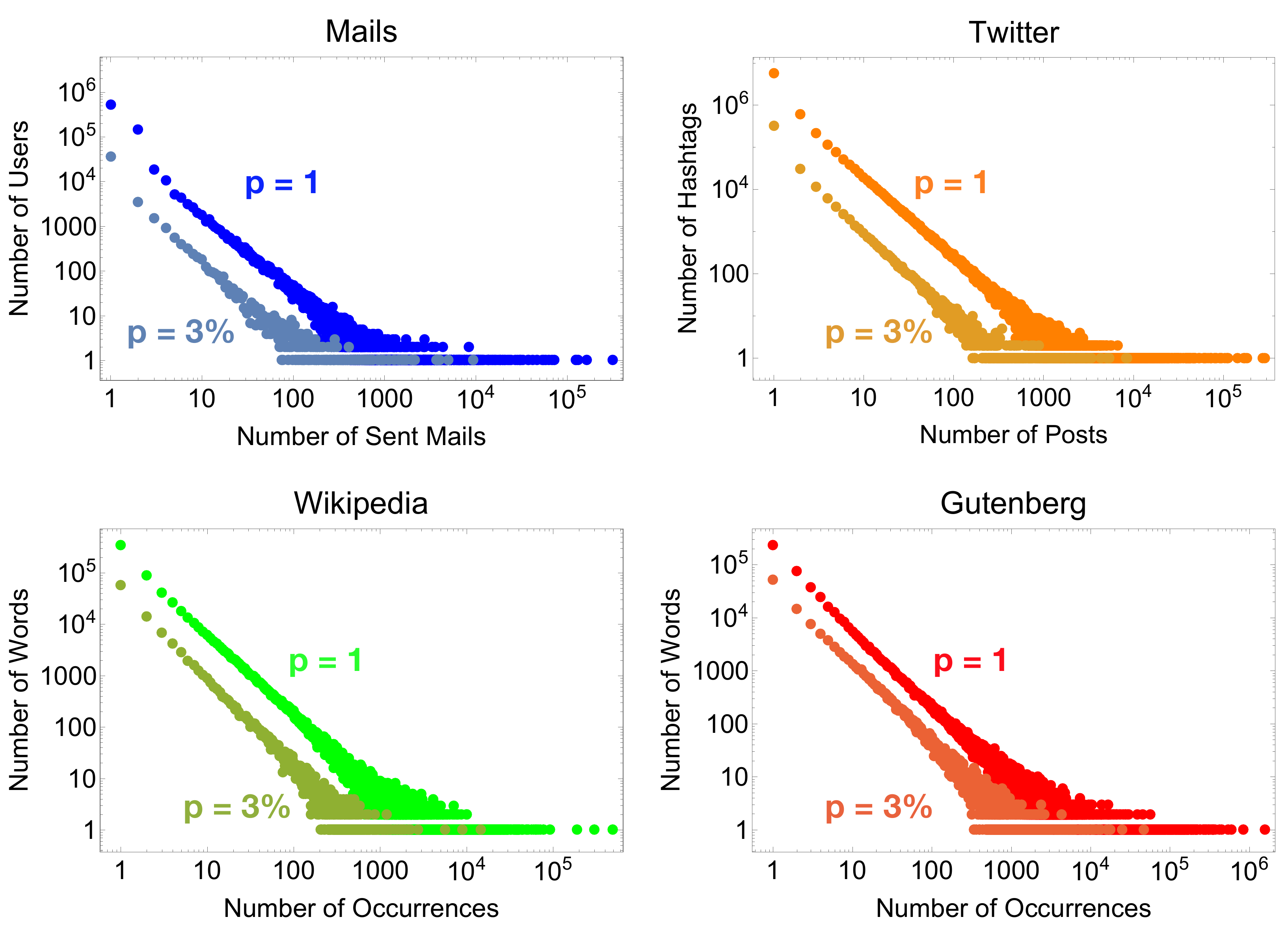}
	\caption{\textbf{Universality and form invariance of the empirical RSAs.}  Empirical RSA curves at global scale ($p=1$) and local scale ($p^*=3\%$) are shown. RSA is scale-free in all the four datasets analyzed, with a power law form maintained through different human activities and different scales. RSA scale-invariance property allows for a successful implementation of our theoretical framework.}\label{RSA-invariant-3}
\end{figure}

\noindent In the main text we showed the results we obtained with our upscaling method when sampling a fraction $p^*=5\%$ of the four databases. We performed the same tests also for a local scale $p^*=3\%$, with similar results.\\
First of all, as shown in \figurename~\ref{RSA-invariant-3}, also for the case $p^*=3\%$ we observe the scale invariance property of the empirical RSAs for all the human activity datasets considered.  
\\
Moreover, as for $p^*=5\%,$ we tested the reliability of estimator (\ref{S}) in predicting the total number of species in the social networks when only a random portion of them is extracted. \tablename~\ref{table-upscaling-3} displays the relative percentage error we obtained for the different databases together with the total dataset composition and the values of the parameters fitted from the empirical RSAs at $p^*=3\%$.
\begin{table}
	\centering
	\begin{tabular}{ccccc}
		\hline
		& Emails & Twitter & Wikipedia & Gutenberg\\
		\hline
		Species & $780,142$ & $6,972,453$ & $673, 872$ & $554,193$\\
		Individuals & $6,914,872$ & $34,696,973$ & $29,606,116$ & $126,289,661$ \\
		$\mathbf{r}$ & $-0.788$ & $ -0.828$ & $-0.549$ & $-0.422$\\
		$\mathbf{\xi_{p^*}}$ & $0.9997$ & $0.9976$ & $0.9987$ & $0.9994$\\
		Relative Error &  -2.74\% & 4.41\% & 8.22\% & -3.52\%\\
		\hline
	\end{tabular}
	\caption{\textbf{Predicted relative errors.} Upscaling results for the number of species of the four analysed datasets from a local sample covering a fraction $p^*=3\%$ of the global database. For each database, we display the number of species (users, hashtags, words)  and individuals (sent mails, posts, occurrences) at the global scale, together with the fitted RSA parameters at the sampled scale and the relative percentage error between the true number of species and the one predicted by our framework.}\label{table-upscaling-3}
\end{table}

\subsection{Upscaling results for popularity change}
\noindent In the main text we exhibited in Table 2 the results for the predictions of popularity (via the conditional estimator \ref{eq:CondEst}) in the unsurveyed fraction $1-p^*=0.95$ of the population for a fixed value of the local popularity threshold $L=10$. In \tablename~\ref{table-upscaling-popularity} we show the results obtained for different values of $L$ and $K$.
\begin{table}
	\centering
	\caption{\textbf{Percentage errors for popularity change predictions in Twitter database.} 
		For different values of $L=10,\ 40,\ 55$ and different values of $K$(first and second column), we estimated, from ten different Twitter sub-samples, the number of species having abundance at least $K$ at the unobserved scale $1-p^*=95\%$ given that they have abundance at least $L$ at the sampled scale $p^*=5\%$ (see estimator 4 of the main text).
		The average true number of species $S_{1-p^*}(\geq K| \geq L)$ and the average one predicted by our method among the ten sub-samples are displayed in the third and fourth columns. Finally, in the last two columns, we inserted the mean and the variance of the relative error obtained in the ten predictions.}\label{table-upscaling-popularity}
	\medskip
	\medskip
	\begin{tabular}{cccccc} 
		\hline
		$L$ & $K$ & $S_{1-p^*}(\geq K| \geq L)$ & $\hat{S}_{1-p^*}(\geq K| \geq L)$ & Relative Error & Variance \\
		\hline
		$10$ & $77$ & $14266$ & $14274.38$  & $-0.0029$ & $0.0012$ \\
		$10$ & $115$ & $14113$ & $14105.65$  & $0.0534$ & $0.0151$ \\
		$10$ & $154$ & $13551$ & $13544.76$  & $0.2457$ & $0.0428$ \\
		$10$ & $192$ & $12509$ & $12584.32$  & $0.4679$ & $0.0731$ \\
		$10$ & $231$ & $11305$ & $11366.66$  & $0.5372$ & $0.0965$ \\
		\hline
		$40$ & $362$ & $3749$ & $3748.99$ & $-0.0001$ & $\approx 0$ \\
		$40$ & $543$ & $3742$ & $3741.96$ & $0.0393$ & $0.0058$ \\
		$40$ & $724$ & $3591$ & $3578.83$ & $-0.0715$ & $0.0668$ \\
		$40$ & $905$ & $3096$ & $3091.45$ & $0.0368$ & $0.0660$ \\
		$40$ & $1086$ & $2600$ & $2582.75$ & $-0.5634$ & $0.0370$ \\
		\hline
		$55$ & $504$ & $2673$ & $2673.00$ & $\approx 0$ & $\approx 0$ \\
		$55$ & $756$ & $2672$ & $2670.96$ & $-0.0141$ & $0.0013$ \\
		$55$ & $1008$ & $2569$ & $2567.71$ & $-0.0978$ & $0.0565$ \\
		$55$ & $1260$ & $2195$ & $2199.11$ & $0.0023$ & $0.0557$ \\
		$55$ & $1512$ & $1806$ & $1820.01$ & $0.1286$ & $0.2070$ \\
		\hline
	\end{tabular}
\end{table} 

\subsection{Local Analysis}
\begin{figure}[h]
	\caption{\textbf{Relative percentage errors at different sub-scales from $p^*=5\%$.} Starting from a sub-sample at $p^*=5\%$ of each human activity database, we sub-sampled it at different spatial sub-scales $p^{**}\in\{10\%,\dots,90\%\}$ and computed the relative percentage error between the number of predicted species, $\hat{S}^*$, and the true number of species, $S^*$, observed in the sample at $p^*$, here considered as the global scale ($p^*=1$.)} \label{local-analysis}
	\includegraphics[width=\textwidth]{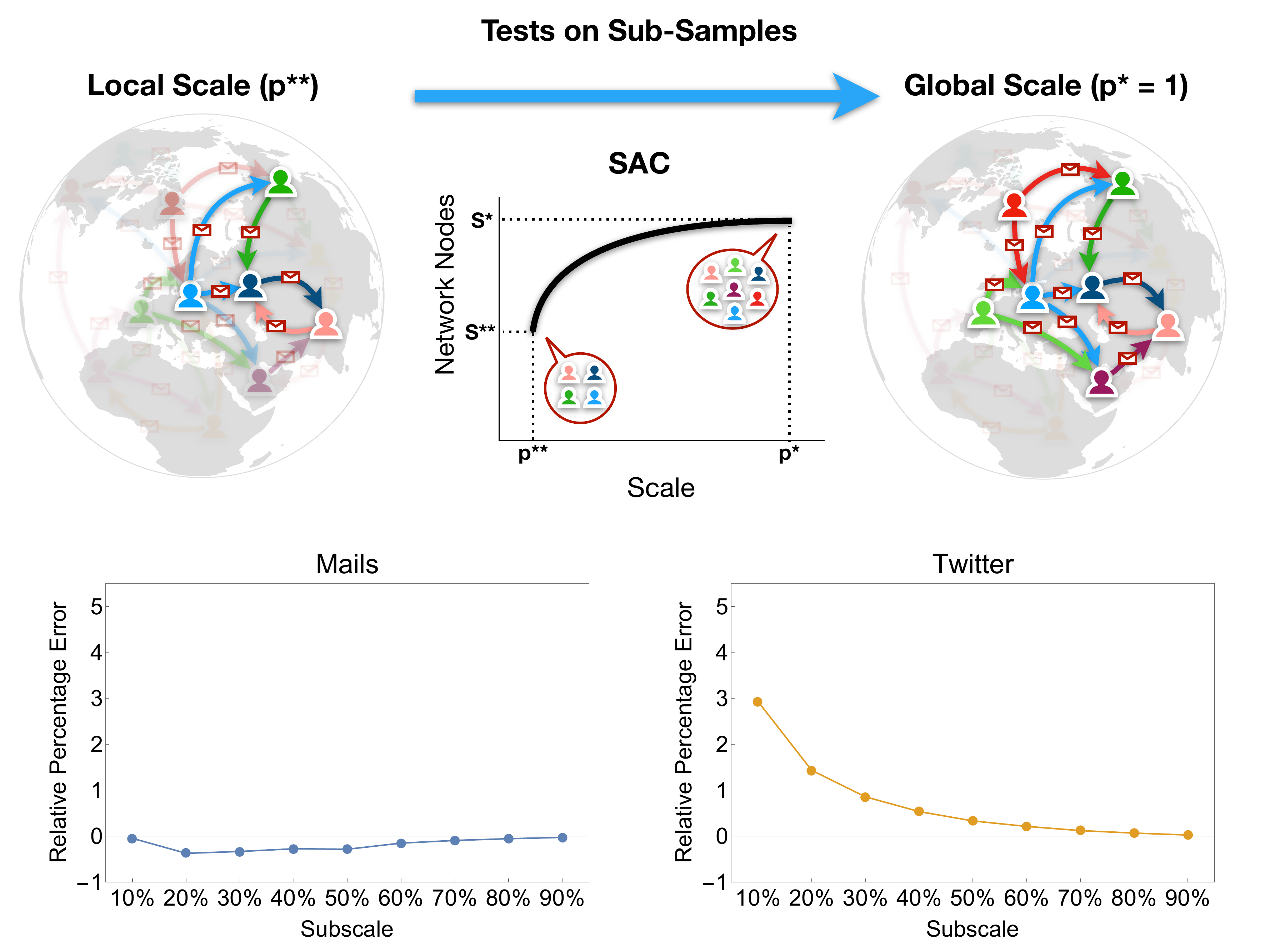}
	\includegraphics[width=\textwidth]{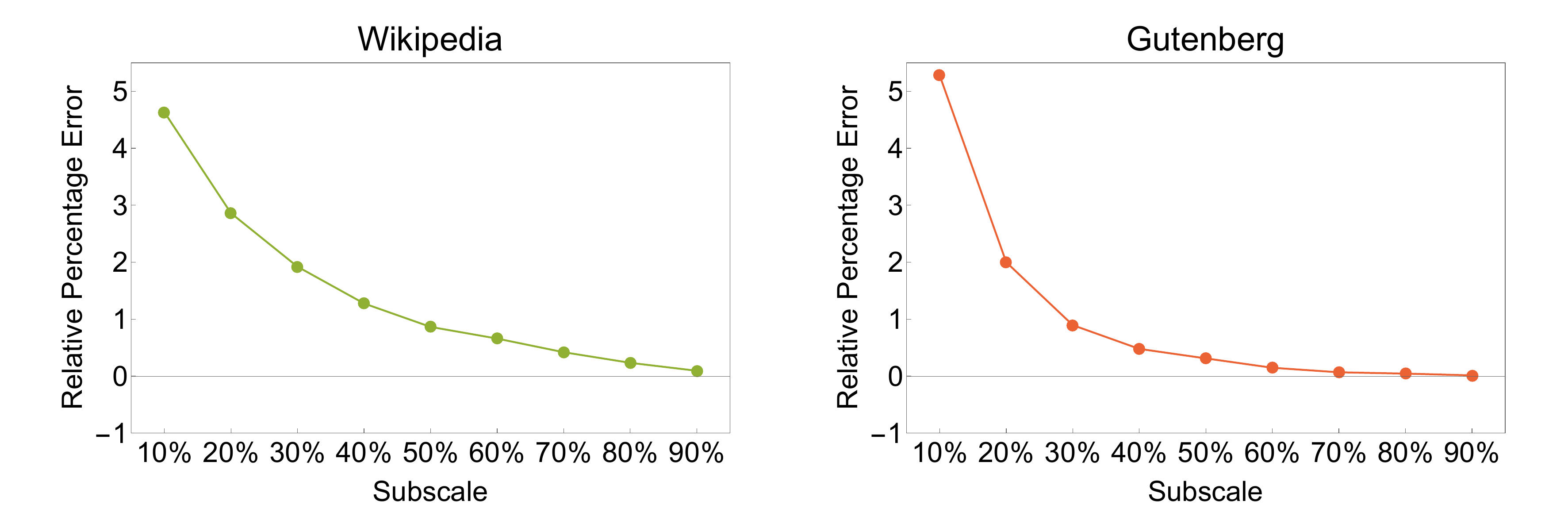}
\end{figure}

\begin{figure}[h]
	\caption{\textbf{Relative percentage errors at different sub-scales from $p^*=3\%$.} Starting from a sub-sample at $p^*=3\%$ of each human activity database, we sub-sampled it at different spatial sub-scales $p^{**}\in\{10\%,\dots,90\%\}$ and computed the relative percentage error between the number of predicted species, $\hat{S}^*$, and the true number of species, $S^*$, observed in the sample at $p^*$, here considered as the global scale ($p^*=1$.)}\label{local-analysis-3}
	\includegraphics[width=\textwidth]{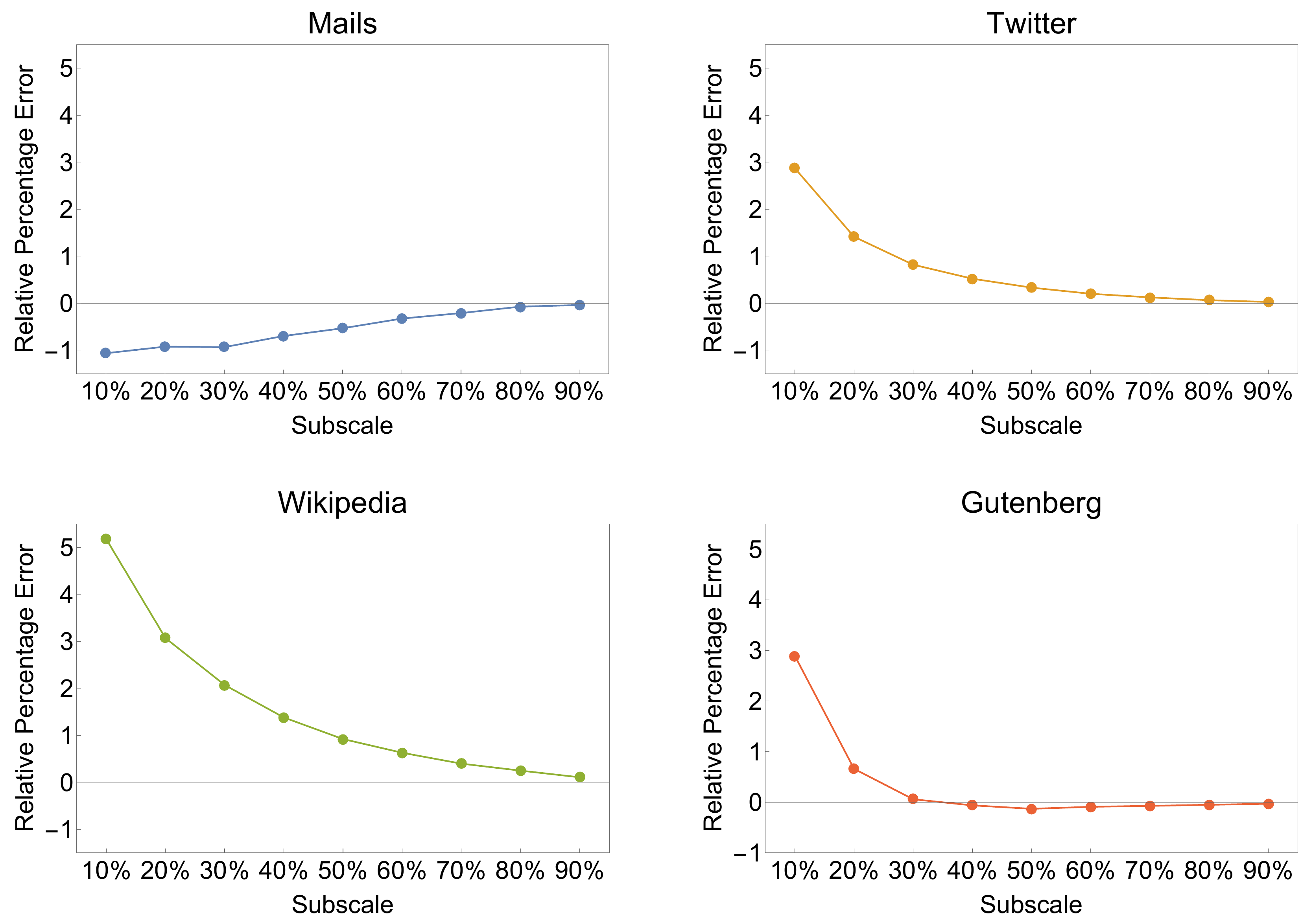}
\end{figure}

\noindent We also tested how estimator (\ref{S}) performs on different spatial sub-scales. In this case, due to the huge amount of data, we chose to work with a smaller datasets for a systematic analysis. In particular, we consider as global four sub-samples of the original datasets each covering a fraction $p^*=5\%$ of the total amount of data (see \figurename~\ref{local-analysis}).\\
We then randomly sub-sampled the smaller resulting database at different scales $p^{**}$ ranging from 10\% to 90\% and applied our framework to predict the number of species observed at $p^*$.\\
In \figurename~\ref{local-analysis}, bottom panels, we displayed the relative percentage error graphs between the predicted number of species, $\hat{S}^*$, and the true number, $S^*$, from local information at the different sub-scales $p^{**}$. We see that, for all datasets and sub-scales, our method always led to an error below 5\%. Moreover, it displays an intuitive decreasing behavior as the available information increases, a desirable property for an estimator.
We performed the same analysis also starting from a sample at the scale $p^*=3\%$, obtaining comparable results (see \figurename~\ref{local-analysis-3}). 

\clearpage



\section*{Acknowledgements}
A.T. acknowledges financial support  from \textit{neXt} grant, Department of Mathematics \textquotedblleft Tullio Levi-Civita" of University of Padova. S. Suweis and A.T. acknowledge STARS grant 2019 from University of Padova. S. Stivanello acknowledges financial support from Progetto Dottorati - Fondazione Cassa di Risparmio di Padova e Rovigo. A.M. was supported by Excellence Project 2017 of the Cariparo Foundation.  Stefano Favaro received funding from the European Research Council (ERC) under the European Union's Horizon 2020 research and innovation programme under grant agreement No 817257. Stefano Favaro gratefully acknowledge the financial support from the Italian Ministry of Education, University and Research (MIUR), ``Dipartimenti di Eccellenza" grant 2018-2022.
	
\section*{Author Contributions} A.T, S.Stivanello, A.M., S.Suweis, S.F. and M.F. designed research, performed research, analysed data and worte the paper.
	
\section*{Author Information} The authors declare no competing financial interests. Readers are welcome to comment on the online version of the paper. Correspondence and requests for materials should be addressed to M.F. (marco.formentin@unipd.it) or to A.T. (anna.tovo@unipd.it).

\bibliographystyle{plain}
\bibliography{sample}

\end{document}